\documentclass[conference]{IEEEtran}

  \usepackage{caption}
  \usepackage{comment}
  \usepackage{lipsum}
\usepackage{cite}
\usepackage{amsmath,amssymb,amsfonts}
\usepackage{algorithmic}
\usepackage[applemac]{inputenc} 
\usepackage{graphicx}
\newcommand{\RN}[1]{%
  \textup{\uppercase\expandafter{\romannumeral#1}}%
}
\newcommand*\sepline{%
   \begin{center}
     \rule[1ex]{1\textwidth}{.5pt}
   \end{center}}
\usepackage{float}
\usepackage{cuted,nccmath}
\usepackage{flushend}
 \usepackage{booktabs}
 \usepackage{multirow}
 \usepackage[table,xcdraw]{xcolor}
\usepackage{textcomp}
\usepackage{mathtools, nccmath}
\usepackage{xcolor}
\def\BibTeX{{\rm B\kern-.05em{\sc i\kern-.025em b}\kern-.08em
    T\kern-.1667em\lower.7ex\hbox{E}\kern-.125emX}}
\begin{document}

\title{Physical Layer Security for V2I Communications:  Reflecting Surfaces  Vs.  Relaying
{\footnotesize \textsuperscript{ }}
\thanks{}
}

\author{\IEEEauthorblockN{\textsuperscript{} Neji Mensi and Danda B. Rawat}
\IEEEauthorblockA{\textit{Department of Electrical Engineering and Computer Science} \\
\textit{Howard University}\\
Washington, DC 20059, USA \\
neji.mensi@bison.howard.edu, danda.rawat@howard.edu }
\and
\IEEEauthorblockN{\textsuperscript{} Elyes Balti}
\IEEEauthorblockA{\textit{Wireless Networking and Communications Group} \\
\textit{The University of Texas at Austin}\\
Austin, TX 78712, USA\\
ebalti@utexas.edu}

}

\maketitle

\begin{abstract}
Wireless vehicular network (WVN) is exponentially gaining attention  from industries and researchers since it is the Keystone of intelligent transportation systems (ITS). Despite the sophisticated features and services that it can offer, it is susceptible to networking attacks such as eavesdropping threats where the confidential transmitted signal could be overheard by a malicious entity. In this paper, we intend to study the physical layer security (PLS) where we consider the eavesdropping attack for vehicle-to-infrastructure (V2I) communications. We analyze the average secrecy capacity, under different scenarios by comparing the performances of employing the decode-and-forward  (D relay, the amplify-and-forward fixed gain (AFFG) relay, and the intelligent reflecting surface (IRS). Actually, this comparison investigates the efficiency of IRS comparing to the traditional relaying systems, since it was introduced as a novel paradigm in wireless technology with highly promising potential, especially in 5G and 6G.

\end{abstract}


\section{Introduction}
Intelligent transportation systems (ITS) has been growing tremendously in term of new sophisticated technologies and capabilities. In particular, wireless vehicular network (WVN) is the spine of ITS where it provides a variety of services such as real-time warning messages \cite{realTime,MeScore},   network as a service (NaaS), storage as a service (STaaS), collaboration as a Service (CaaS) \cite{MeTrafficJam,eVCloud}, etc. To reach such goal expectations, WVN should be based on high throughput infrastructures by employing Millimeter Wave (mmWave) \cite{eJEns_Mmwave, mmw,mmwArx}, terahertz (THz), 5G, machine learning, etc. Another major factor that impacts the WVN performances is the concept of security, where ITS should guarantee minimum security criteria to protect the communication links between the different WVN entities.\\ The eavesdropping attack is a passive threat that could negatively influence the offered quality of service (QoS) from a security point of view, where the attacker overhears the channel between the communicating entities and discloses the transmit secret messages. What makes this kind of vulnerability very critical is the fact that it is very challenging to detect an occurring eavesdropping attack.  Therefore, to \let\thefootnote\relax\footnotetext{  This work was supported in part by the U.S. National Science Foundation (NSF) under the grant CNS-1650831.}deal with such menace, several research papers are focusing on physical layer security (PLS) where they intend to increase  the average secrecy capacity between the legitimate entities and the eavesdroppers, increasing the outage probability and the symbol error probability at the attackers ends \cite{SecOut}.\\ Many research attempts discussed PLS by employing relaying in the communication systems. Other suggest the use of the intelligent reflecting surface (IRS) \cite{PLS_RIS}. In fact, the IRS is considered as a trend in wireless telecommunications due to its low power consumption, its promising role to deal with mmWaves/THz blockage by offering multipath signals paths to the destination \cite{RIS_6G2,RIS_6G1}. The IRS was compared to relaying in \cite{IRSvsRelayEmil,IRSvsRelay} with respect to rate performances and power consumption, but according to the best knowledge of the authors, the comparison from a security point of view in the context of eavesdropping attack was not studied yet. \\
In our paper, we are going to analyse the average secrecy capacity while comparing the efficiency of employing an IRS with both relaying systems: relay decode-and-forward (DF), and amplify-and-forward with fixed gain (AFFG).\\
The paper is organized as follows: Section \RN{2} introduces the IRS system model, the attack scenario, and the ergodic capacity expression at the legitimate and the attacker. Section \RN{3} studies the relaying system model and presents closed-form expressions for the average secrecy capacity of DF and AFFG relays. Then, Section \RN{4}   evaluates the performances of the three system scenarios based on the numerical results. Lastly, we outline our conclusion in section \RN{5}.
\section{System Model Using IRS}
\subsection{Attack Description and Channel Model}
As shown in Fig. 1, the base station (BS) noted by $T$ is communicating with the legitimate vehicle in black via the IRS as an intermediate link. Unfortunately, an eavesdropping is listening to the channel and intends to disclose the confidential transmitting information. 
The studied scenario has three channel links: BS-to-IRS, IRS-to-receiver, and IRS-to-eavesdropper.  We assume that the aforementioned links are subject to Nakagami-m fading.

\begin{figure}[]
\begin{center}
\includegraphics[height=50mm, width=0.7\linewidth]{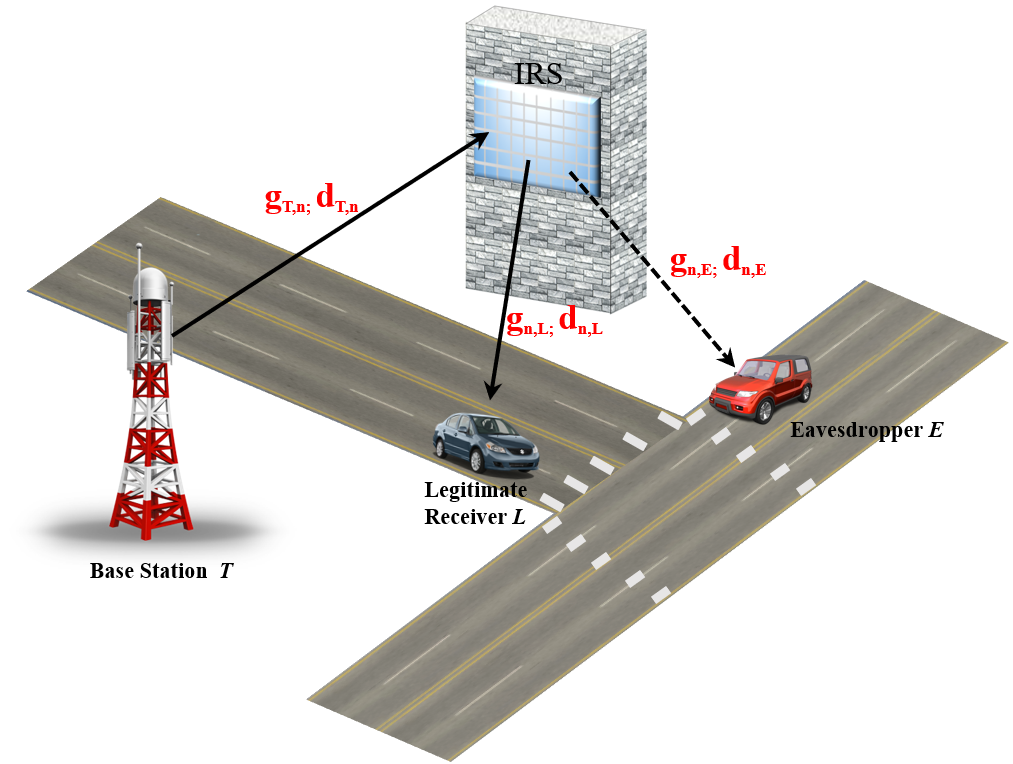}
\caption{V2I communications using IRS in the presence of an eavesdropper.  }
\end{center}
\end{figure}

The received signal at \textit{R} and \textit{E}  are respectively:

\begin{equation}
y_{_i} =\left(\sum_{n=1}^{N}{ h}_{_{T,n}}{h}_{_{n,i}}e^{j\phi_{_n}}\right)x +w_{_i},
\end{equation}
where  $i\in \{L,E\} $, \textit{N} denotes the IRS's elements numbers, $x$ is the transmitted signal by $T$ with power $P_{s}$,  $w_{_i}$ is the additive white Gaussian noise (AWGN) at the receiving node  $i$. The terms ${ h}_{_{T,n}}=\sqrt{d_{_{T,n}}^{-\zeta}}g_{_{_{T,n}}}e^{-j\theta_{n}}$  presents the channel coefficients between $T$ and the $n$-th reflecting element with distance $d_{_{T,n}}$ and channel phase $\theta_n$. ${ h}_{_{n,i}}=\sqrt{d_{_{n,i}}^{-\zeta}}g_{_{_{n,i}}}e^{-j\psi{_n}}$ is the channel coefficients relying the $n$-th reflecting element to the node  $i$ with distance $d_{_{n,i}}$ and channel phase $\psi_n$. The adjustable phase induced by the \textit{n}-th reflecting element is denoted by $\phi_n$ and the pathloss has an exponent $\delta$. \\
The channels fading are assumed to follow Nakagami-m distribution. Hence, $|g_{_{T,n}}|^2$ $\sim$ $\textit{\textsf{G}}(\alpha_{_{T,n}},\beta_{_{T,n}})$ and $|g_{n,i}|^2$ $\sim$ $\textit{\textsf{G}}(\alpha_{_{n,i}},\beta_{_{n,i}})$, where
$\alpha_{_{T,n}}$, $\alpha_{_{n,i}}$ are the corresponding scale parameters and $\beta{_{_{T,n}}}$, $\beta{_{n,i}}$ are the shape parameters. Therefore, $|g_{_{T,n}}g_{_{n,i}}|^{2}$  follows Gamma-Gamma distribution  $\textit{\textsf{GG}} (\alpha{_{i_n}},\beta{_{i_n}})$, where $\alpha{_{i_n}}=\frac{\alpha_{_{T,n}}+\alpha_{_{n,i}}}{2}$ and $\beta{_{i_n}}=\beta_{_{T,n}}\beta_{_{n,i}}$.
The signal-to-noise ratio (SNR) at the legitimate receiver $L$ and at the eavesdropper $E$ is expressed as follows:
\begin{equation}
\begin{aligned}
{ \gamma_{_i}} &=\frac{\sum_{n=1}^{N}  P_{_S}|h_{_{T,n}}h_{_{n,i}}|^{2} }{w_{_i}}\\ &=\frac{\sum_{n=1}^{N}  P_{_S}|g_{_{T,n}}g_{_{n,i}}e^{j(\phi{_n}-\psi{_n}-\theta_{_n})}|^{2} d_{_{T,n}}^{-\zeta}d_{_{n,i}}^{-\zeta} }{w_{_i}}
=\sum_{n=1}^{N} \gamma_{_{i_n}},
\end{aligned}
\end{equation}
where $ \gamma_{_{i_n}}$ $\sim$ $\textit{\textsf{GG}} (\alpha{_{i_n}},\beta{_{i_n}})$ is the overall SNR from $T$ to the final node $i$ reflecting by the $n$-th element.
We assume that the IRS has a perfect knowledge of the channel. Hence, we choose $\phi_n = \psi{_n}+\theta_{_n}$ to maximize the transmitted signal.\\
The following probability density function (PDF) of $\gamma_{_{i_n}}$ is expressed as follows:  
\begin{equation}
\begin{aligned}
{ f}_{\gamma_{_{i_n}}}(\gamma) =\frac{2(\beta{_{i_n}})^{\alpha{_{i_n}}}\gamma^{\alpha{_{i_n}}-1}}{\Gamma{(\alpha_{_{T,n}})}\Gamma{(\alpha_{_{n,i}}})}K_{v_{i_n}}\left(2\sqrt{\gamma\beta{_{i_n}}}\right),
\end{aligned}
\end{equation}
where $v_{i_n}=\alpha_{_{T,n}}-\alpha_{_{n,i}}$ and $K(\cdot)$ is the modified Bessel function of the second kind and order $v_{i_n}$.
\subsection{Average Secrecy Capacity Analysis}
The general expression of the ergodic capacity of $\gamma_{_i}$ is  defined by:
\begin{equation}
   \overline{C}= \mathbb{E}\left[\log_{2}(1+\gamma_{_i})\right].
\end{equation}
Then, the average secrecy capacity $\overline{C}_{_S}$ can be expressed by:
\begin{equation}
\overline{C}_{s} = \max(\overline{C}_{_L}-\overline{C}_{_{E}},0) 
\end{equation}
where $\overline{C}_{_L}$ and $\overline{C}_{_E}$ are the ergodic capacity at the legitimate receiver $L$ and the attacker $E$, respectively. \\
To compute $\overline{{C}}_{{s}}$, we recall the identity [\citenum{MRC}, Eq. (6)]
\begin{equation}
\ln(1+\eta) = \int\limits_{0}^{\infty} \frac{1}{z}  (1-e^{-\eta z})dz.
\end{equation}
Then, by substituting $\gamma_{_{_i}}=\sum_{n=1}^{N} \gamma_{_{i_n}}$ in Eq. (24), we obtain the expression of $\overline{{C}}_{{s}}$ as follows:
\begin{equation}
\overline{{C}}_{{s}} = \frac{1}{\ln(2)} \int\limits_{0}^{\infty} \frac{1}{z}  (1-\mathcal{M}_{\gamma_{_i}}(z))e^{-z}dz,
\end{equation}
where  $\mathcal{M}_{\gamma_{_i}}(\cdot) $  is the moment generation function (MGF) of $\gamma_{_i}$, which can be derived as:
\begin{equation}
\begin{aligned}
\mathcal{M}_{\gamma_{_i}}(z)&=\mathbb{E}\left[ e^{-z\gamma_{_{_i}}}\right]
=\mathbb{E}\left[ e^{-z \sum_{n=1}^{N}\gamma_{_{i_n}}}\right]=\prod_{n=1}^{N}\mathbb{E}\left[ e^{-z\gamma_{_{i_n}}}\right].
\end{aligned}
\end{equation}
To facilitate the MGF derivation, we can replace the Bessel function in Eq. (3) as follows \cite{BesselToMeijer}:
\begin{equation}
\begin{aligned}
    K_{v_{i_n}}\left(2\sqrt{\gamma\beta{_{i_n}}}\right)=\frac{1}{2}G_{0,2}^{2,0} 
\left(  \gamma\beta{_{i_n}}
                          \bigg| \begin{matrix} -\\
                                       \frac{v_{i_n}}{2}, \frac{-v_{i_n}}{2} \end{matrix}
\right),
   \end{aligned}
\end{equation}
where $G^{m,n}_{p,q}(\cdot)$ is the Meijer G-Function. Then, by using Eqs. (3, 8, and 9) and [\citenum{Tab}, Eq (7.813.1)], the MGF becomes:

\begin{equation}
\begin{aligned}
\mathcal{M}_{\gamma_{_i}}(z)&=\prod_{n=1}^{N} \left[\frac{(\beta{_{i_n}})^{\alpha{_{i_n}}}z^{-\alpha{_{i_n}}}}{\Gamma{(\alpha_{_{T,n}})}\Gamma{(\alpha_{_{n,i}}})}G_{1,2}^{2,1} \left(  \frac{\beta{_{i_n}}}{z}
                          \bigg| \begin{matrix} 1-\alpha{_{i_n}}\\
                                       \frac{v_{i_n}}{2}, \frac{-v_{i_n}}{2} \end{matrix}
\right) \right]\\
&= \left[\frac{(\beta{_{i}})^{\alpha{_{i}}}z^{-\alpha{_{i}}}}{\Gamma{(\alpha_{_{T,S}})}\Gamma{(\alpha_{_{S,i}}})}G_{1,2}^{2,1} \left(  \frac{\beta_{_{i}}}{z}
                          \bigg| \begin{matrix} 1-\alpha_{_{i}}\\
                                       \frac{v_i}{2}, \frac{-v_i}{2} \end{matrix}
\right) \right]^N,
\end{aligned}
\end{equation}
where we assumed that the distance between each reflecting element $n\in \{1,...,N\}$ is negligible, which implies that the fading parameters are equal for all $N$ channels. Hence, $\beta{_{i_n}}=\beta_{_{i}}$, $\alpha{_{i_n}}=\alpha{_i}$, $\Gamma{(\alpha_{_{T,n}})}=\Gamma{(\alpha_{_{T,S}})}$ (the couple "T,S" refers to Transmitter-to-Surface), and $\Gamma{(\alpha_{_{n,i}})}=\Gamma{(\alpha_{_{S,i}})}$  (the couple "S, i" refers to Surface-to-receiver $i$).\\
Now, we substitute Eq. (10) in Eq. (7) to obtain the ergodic capacity at $i$ as follows:

\begin{equation}
\begin{aligned}
\overline{{C}}_{{_i}} = \frac{1}{\ln(2)} \int\limits_{0}^{\infty}& \frac{1}{z}  \left(1-\left[\frac{(\beta_{_{i}})^{\alpha_{_{i}}}z^{-\alpha_{_{i}}}}{\Gamma{(\alpha_{_{T,S}})}\Gamma{(\alpha_{_{S,i}}})}\right.\right. \\ &\left.\left. \times G_{1,2}^{2,1} \left(  \frac{\beta_{_{i}}}{z}
                          \bigg| \begin{matrix} 1-\alpha_{_{i}}\\
                                       \frac{v_i}{2}, \frac{-v_i}{2} \end{matrix}
\right) \right]^N\right)e^{-z}dz,
\end{aligned}
\end{equation}
At this step, we derive the final expression of the average secrecy capacity $C_{_S}$ by substituting Eq. (11) in Eq. (5), which is presented by Eq. (12) at the top of the next page.
\begin{figure*}
\begin{equation}
\begin{aligned}
\overline{{C}}_{{_S}}^{_{IRS }} = \frac{1}{\ln(2)} \int\limits_{0}^{\infty} \frac{e^{-z}}{z}  \left\{\left( \left.\left.1-\left[\frac{(\beta_{_{L}})^{\alpha_{_{L}}}(z)^{-\alpha_{_{L}}}}{\Gamma{(\alpha_{_{T,S}})}\Gamma{(\alpha_{_{S,L}}})} \right.\right.\right.\right.\right.&\left.\left.\left.G_{1,2}^{2,1} \left(  \frac{\beta_{{L}}}{z}
                          \bigg| \begin{matrix} 1-\alpha_{_{L}}\\
                                       \frac{v_{L}}{2}, \frac{-v_{L}}{2} \end{matrix}
\right) \right]^N\right)\right.\\&-\left. \left(1-\left[\frac{(\beta_{_{E}})^{\alpha_{_{E}}}(z)^{-\alpha_{_{E}}}}{\Gamma{(\alpha_{_{T,S}})}\Gamma{(\alpha_{_{S,E}}})} G_{1,2}^{2,1} \left(  \frac{\beta_{_{E}}}{z}
                          \bigg| \begin{matrix} 1-\alpha_{_{E}}\\
                                       \frac{v_{_E}}{2}, \frac{-v_{_E}}{2} \end{matrix}
\right) \right]^N\right)\right\}dz.
\end{aligned}
\end{equation}
\sepline
\end{figure*}

\section{System Model Using Relay }
\subsection{Attack Description and Channel Model}
\begin{figure}[H]
\begin{center}
\includegraphics[ height=50mm, width=0.7\linewidth]{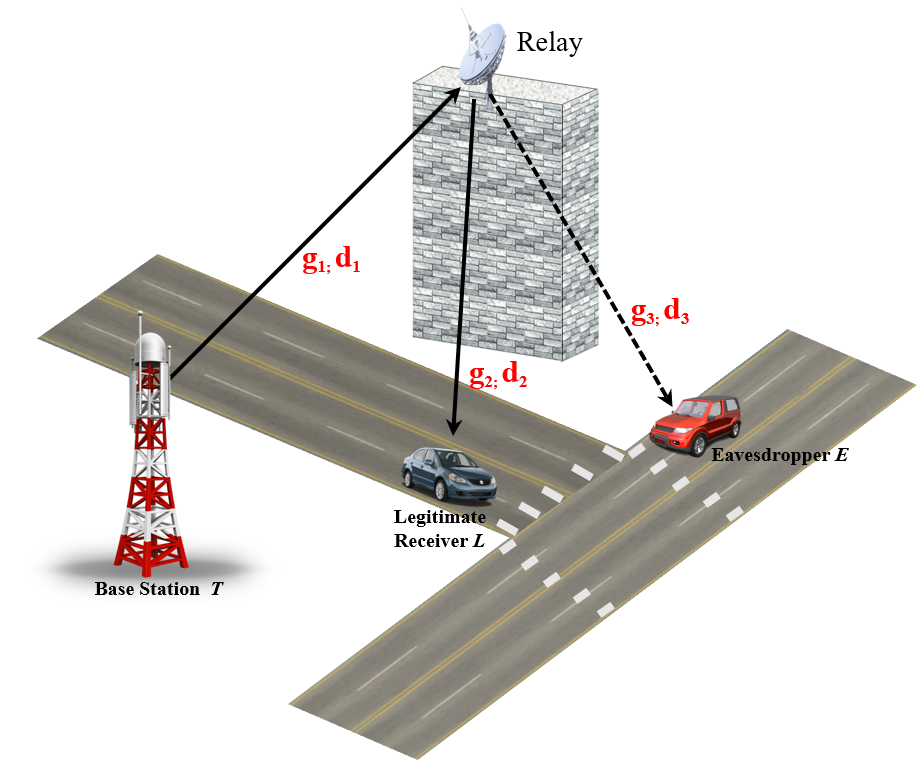}
\caption{V2I communications using Relay in the presence of an eavesdropper.  }
\end{center}
\end{figure}
In this section, we are considering the analysis of the relay security performances as shown in Fig. 2, where we examine the same eavesdropping scenario as described in the IRS's Section. We note by $h_{{1}}$ the channel coefficient of the link  from the transmitter $T$ to the relay $R$,  $h_{{2}}$ denotes the channel  coefficient of the link from $R$ to the legitimate receiver $L$, and $h_{_{3}}$ describing the hop-link between $R$ and the eavesdropper $E$.  We assume that the aforementioned links are subject to Nakagami-m distribution.

Assuming AFFG relaying, the received signal at $L$ and $E$ are respectively:
\begin{equation}
\begin{aligned}
y_{_L} &=Gh_{_{2}}(h_{_{1}}x + w_{_R}) +w_{_L}\\
&=G\sqrt{d_{_{2}}^{-\zeta}}g_{_{{2}}}\left(\sqrt{d_{_{1}}^{-\zeta}}g_{_{{1}}}x+w_{_R}\right) +w_{_L},
\end{aligned}
\end{equation}
\begin{equation}
\begin{aligned}
y_{_E} &=G h_{_{3}}(h_{_{1}}x + w_{_R}) +w_{_E}\\
&=G\sqrt{d_{_{3}}^{-\zeta}}g_{_{{3}}}\left(\sqrt{d_{_{1}}^{-\zeta}}g_{_{{1}}}x+w_{_R}\right) +w_{_E},
\end{aligned}
\end{equation}

where G is the amplification gain, $x$ is the transmitted signal by $T$ with power $P_{s}$, $w_{_R}$, $w_{_L}$, and $w_{_E}$   are the respectively AWGN at $R$, $L$, and $E$. The distances $T-to-R$, $R-to-L$, and $R-to-E$ are respectively, $d_{_{1}}$, $d_{_{2}}$, and $d_{_{3}}$. The path loss is described by the exponent $\delta$.
The channels fading are assumed to follow Nakagami-m distribution. Hence, $|g_{_1}|^2$ follows Gamma distribution  $\textit{\textsf{G}}(\alpha_{_{_{1}}},\beta_{_{_{1}}})$, $|g_{_2}|^2$ $\sim$ $\textit{\textsf{G}}(\alpha_{_{2}},\beta_{_{_{2}}})$, and $|g_{_3}|^2$ $\sim$ $\textit{\textsf{G}}(\alpha_{_{3}},\beta_{_{_{3}}})$ where
$\alpha_{_{_1}}$, $\alpha_{_{_2}}$, and $\alpha_{_{_3}}$ are the corresponding scale parameters and $\beta{_{_{1}}}$, $\beta{_{_{2}}}$ and $\beta{_{_{3}}}$ are the shape parameters. Therefore, each hop: $T-to-R$, $R-to-i$ has the PDF and cumulative distribution function (CDF):
\begin{equation}
\begin{aligned}
{ f}_{\gamma_{_{a}}}(\gamma) =\frac{\beta_{_a}^{\alpha_{_a}}\gamma^{\left(\alpha_{_a}-1\right)}\exp({-\beta_{_a}\gamma})}{\Gamma(\alpha_{_a})},
\end{aligned}
\end{equation}
\begin{equation}
\begin{aligned}
{ F}_{\gamma_{_a}}(\gamma) &= P[\gamma_{_a}\leq\gamma]  =1-\frac{\Gamma(\alpha_{_a},\beta_{_a}\gamma) }{\Gamma(\alpha_{_a})}, 
\end{aligned}
\end{equation}
where $a\in\{1,2,3\}$.
For the sake of derivation simplicity, we assume that $\alpha{_a}$ is a positive integer. Accordingly, the CDF has the subsequent series expansion:
\begin{equation}
\begin{aligned}
{ F}_{\gamma_{_a}}(\gamma) =1- \sum_{j=0}^{\alpha_{a}-1} \frac{(\beta_{_a}\gamma)^{j}}{j!}e^{-\beta_{_a}\gamma} .
\end{aligned}
\end{equation}

After presenting the channel model, we are going to study the average secrecy capacity for DF and AFFG. For the sake of organization, and before starting the study of each type of relay, we will present the used formula to compute the ergodic capacity at the final receiving node. It could be derived after applying the integration by part on Eq. (4):
\begin{equation}
\begin{aligned}
   \overline{C}=\frac{1}{\ln(2)}\int_{0}^{\infty} \frac{\overline{F}_{\gamma_{_i}}(\gamma)}{1+\gamma} d\gamma,
\end{aligned}
\end{equation}
where $i\in\{L,E\}$, $\overline{F}_{\gamma_{_i}}(\cdot)$ is the complementary of the CDF. 
\subsection{Average Secrecy Capacity for DF Relay}
The SNR at the node $i$ using a DF relay is expressed as follows\cite{eAppr}:
\begin{equation}
\begin{aligned}
 \gamma^{_{_{DF}}}_{_i} =\min\{\gamma{_{_{1}}},\gamma{_{_{b}}}\},
\end{aligned}
\end{equation}
where $b\in\{2,3\}$.\\
To compute the ergodic capacity, we have to derive first $\overline{F}_{\gamma^{_{_{DF}}}_{_i}}$: 
\begin{equation}
\begin{aligned}
\overline{F}_{\gamma^{_{_{DF}}}_{_i}}(\gamma) &=P[\min(\gamma{_{_{_{1}}}},\gamma{_{_{_{b}}}})>\gamma]\stackrel{\Delta}=P[\gamma{_{_{_{1}}}}>\gamma]P[\gamma{_{_{_{b}}}}>\gamma]\\
&=\sum_{j=0}^{\alpha_{_{_{1}}}-1}\sum_{p=0}^{{\alpha_{_{_{b}}}-1}} \frac{\beta_{_{_{1}}}^{j}\beta_{_{_{b}}}^{p}\gamma^{j+p} }{j!p!}e^{-\gamma(\beta_{_{_{1}}}\beta_{_{_{b}}})},
\end{aligned}
\end{equation}
where the equality passage $\stackrel{\Delta}=$ assumes that  $\gamma_{_{_{1}}}$ and $\gamma_{_{_{b}}}$ are considered to be i.i.d (Independent and identically distributed random variables).
Then, we substitute Eq. (20) in Eq. (18) while performing the following transformation \cite{eTransG}:
\begin{equation}
\frac{1}{1+\gamma}=G_{1,1}^{1,1} 
\left(  \gamma
                          \bigg| \begin{matrix} 0 \\
                                       0 \end{matrix}
\right)
\end{equation}

Finally, by referring to the  integral identity [\citenum{Tab}, Eq (7.813.1)], we obtain: 
\begin{equation}
\begin{aligned}
\overline{C}_{\gamma_{i}}^{_{DF}}(\gamma) &=\frac{1}{\ln(2)}\sum_{j=0}^{\alpha_{_{_{1}}}-1}\sum_{p=0}^{{\alpha_{_{_{b}}}-1}} \frac{\beta_{_{_{1}}}^{j}\beta_{_{_{b}}}^{p} }{j!p!}(\beta_{_{_{1}}}+\beta_{_{_{b}}})^{-(j+p+1)}\\& \times G_{1,2}^{2,1} \left(  \frac{1}{\beta_{_{_{1}}}+\beta_{_{_{b}}}}
                          \bigg| \begin{matrix} -(j+p),0\\
                                       0 \end{matrix}
\right)
\end{aligned}
\end{equation}
Therefore, we can derive the average capacity for DF relay by substituting Eq. (22) in Eq. (5). The expression is given by (23) at the top of the next page.
\begin{figure*}
\begin{equation}
\begin{aligned}
{\overline{C}}_{_S}^{_{DF}}(\gamma) =\frac{1}{\ln(2)}\left\{\sum_{j=0}^{\alpha_{_{_{1}}}-1}\sum_{p=0}^{{\alpha_{_{_{L}}}-1}} \frac{\beta_{_{_{1}}}^{j}\beta_{_{_{L}}}^{p} }{j!p!}\right.&\left.(\beta_{_{_{1}}}+\beta_{_{_{L}}})^{-(j+p+1)} 
G_{1,2}^{2,1} \left(  \frac{1}{\beta_{_{_{1}}}+\beta_{_{_{L}}}}
                          \bigg| \begin{matrix} -(j+p),0\\
                                       0 \end{matrix}
\right)\right.
\\& \left.- \sum_{j=0}^{\alpha_{_{_{1}}}-1}\sum_{p=0}^{{\alpha_{_{_{E}}}-1}} \frac{\beta_{_{_{1}}}^{j}\beta_{_{_{E}}}^{p} }{j!p!}(\beta_{_{_{1}}}+\beta_{_{_{E}}})^{-(j+p+1)} G_{1,2}^{2,1} \left(  \frac{1}{\beta_{_{_{1}}}+\beta_{_{_{E}}}}
                          \bigg| \begin{matrix} -(j+p),0\\
                                       0 \end{matrix}
\right)\right\}
\end{aligned}
\end{equation}
\end{figure*}

\subsection{Average Secrecy Capacity for AFFG Relay}
The SNR at the node $i$ using an AFFG relay is expressed as follows\cite{eRelayFSO2,eRelayFSO3,eRelayFSO1 }:
\begin{equation}
\begin{aligned}
 \gamma_{_{_{i}}}^{_{AF}} =\frac{\gamma{_{_{1}}}\gamma{_{_{_{b}}}}}{\gamma{_{_{_{1}}}}+\overline{\gamma}{_{_{_{1}}}}+1}=\frac{\gamma{_{_{1}}}\gamma{_{_{_{b}}}}}{\gamma{_{_{_{1}}}}+l},
\end{aligned}
\end{equation}
where $l=\overline{\gamma}{_{_{_{1}}}}+1$.
Therefore, the CDF is presented by:
\begin{equation}
\begin{aligned}
F&_{\gamma_{_{_{i}}}^{_{AF}}}(\gamma)
=1-\sum_{j=0}^{\alpha_{_{_{1}}}-1}\sum_{k=0}^{{j}}{j\choose k} \frac{l^k(\beta_{_{_{1}}}\gamma)^{j}(\beta_{_{_{b}}})^{\alpha_{_{_{b}} }}}{j!\Gamma(\alpha_{_{_{b}}})}e^{-\beta_{_{_{1}}}\gamma}\\
&\times \int\limits_{0}^{\infty}\gamma{_{_{b}}}^{(\alpha{_{_{b}}}-k-1)}\exp{\left(\frac{-\beta{_{_{1}}}l\gamma }{\gamma{_{_{b}}}}-\beta{_{_{b}}}\gamma{_{_{b}}}\right)}d\gamma{_{_{b}}}.
\end{aligned}
\end{equation}
To compute the integral, we recall the identity [\citenum{Tab}, Eq (3.471.9)]. Hence, we obtain:
\begin{equation}
\begin{aligned}
\overline{F}_{\gamma_{_{_{i}}}^{_{AF}}}(\gamma)
=&\sum_{j=0}^{\alpha_{_{_{1}}}-1}\sum_{k=0}^{{j}}{j\choose k} \frac{l^k(\beta_{_{_{1}}}\gamma)^{j}(\beta_{_{_{b}}})^{\alpha_{_{_{b}} }}}{j!\Gamma(\alpha_{_{_{b}}})}e^{-\beta_{_{_{1}}}\gamma}\\
&\times 2\left(\frac{\beta_{_{_{1}}}l\gamma }{\beta_{_{_{b}}}}\right)^{\frac{u_{_b}}{2}}K_{u_{_b}}\left(2\sqrt{\gamma\beta_{_{_{1}}}\beta_{_{_{i}}}l}\right),
\end{aligned}
\end{equation}
where $u_{_b}=\alpha_{_{_{b}}}-k$.
To facilitate the integral computation, we transform the following expressions into Fox-H function:
\begin{equation}
\frac{1}{1+\gamma}=H_{1,1}^{1,1} 
\left(  \gamma
                          \bigg| \begin{matrix} (0,1) \\
                                       (0,1) \end{matrix}
\right)~~ ; ~~e^{-\beta_{_{_{1}}}\gamma}=H_{1,0}^{0,1} 
\left(  \beta_{_{_{1}}}\gamma
                          \bigg| \begin{matrix} -\\
                                       (0,1)  \end{matrix}
\right)
\end{equation}
\begin{equation}
\begin{aligned}
    K_{u_{_b}}\left(2\sqrt{\gamma\beta_{_{_{1}}}\beta_{_{_{b}}}}\right)=\frac{1}{2}H_{0,2}^{2,0} 
\left( \gamma l \beta_{_{_{1}}}\beta_{_{_{b}}}
                          \bigg| \begin{matrix} -\\
                                       (\frac{u_{_b}}{2},1) ,(\frac{-u_{_b}}{2},1) \end{matrix}
\right).
   \end{aligned}
\end{equation}
In the next step, we substitute Eqs. (27) and (28) in Eq. (26), then into Eq. (18) and by using \cite{TripFoxInt, eT} we obtain:
 \begin{equation}
\begin{aligned}
 \overline{C}_{\gamma_{i}}^{_{AF}}(\gamma)
&=\frac{1}{\ln(2)}\sum_{j=0}^{\alpha_{_{_{1}}}-1}\sum_{k=0}^{{j}}{j\choose k} \frac{l^{k}(\beta_{_{_{b}}})^{(\alpha_{_{_{b}}})}}{j!\beta_{_{_{b}}}^{\frac{u_{_b}}{2}}\Gamma(\alpha_{_{_{b}}})}\\
&\times H_{0,1;1,1;0,2}^{1,0;1,1;2,0} 
\left(  \frac{1}{\beta_{_{_{R}}}},\beta_{_{_{b}}}l
                          \bigg| \begin{matrix} -(\frac{u_{_b}}{2}+j);(1,1)\\
                                       (-;-) \end{matrix}
                          \bigg|\right.\\&~~~~~~~~~~~~~~~~~~~~~~\left. \begin{matrix} (0,1)\\
                                       (0,1) \end{matrix}  
                           \bigg| \begin{matrix} -\\
                                       (\frac{u_{_i}}{2},1); (\frac{-u_{_i}}{2},1)  \end{matrix}             
\right),
\end{aligned}
\end{equation}
where  $H_{m_{1},n_{1};m_{2},n_{2};m_{3},n_{3}}^{p_{1},q_{1};p_{2},q_{2};p_{3},q_{3}}(-|(\cdot,\cdot))$ is the bivariate Fox H-function\cite{eBivariate_Relay}.
Therefore, the average secrecy capacity could be deduced by substituting Eq. (29) in Eq. (5), as expressed in Eq. (30) at the top of the next pages
\begin{figure*}
\begin{equation}
\begin{aligned}
 \overline{{C}}_{{_S}}^{_{FG }}
=\frac{1}{\ln(2)}&\left\{\sum_{j=0}^{\alpha_{_{_{1}}}-1}\sum_{k=0}^{{j}}{j\choose k} \frac{l^{k}}{j!(\beta_{_{_{1}}})}\left[\frac{\beta_{_{_{L}}}^{\alpha_{_L}}}{\beta_{_{_{L}}}^{\frac{u_L}{2}}\Gamma(\alpha_{_{_{L}}})} H_{0,1;1,1;0,2}^{1,0;1,1;2,0} 
\left(  \frac{1}{\beta_{_{_{R}}}},\beta_{_{_{L}}}l
                          \bigg| \begin{matrix} -(\frac{-u_{_L}}{2}+j);(1,1)\\
                                       (-;-) \end{matrix}
                          \bigg| \begin{matrix} (0,1)\\
                                       (0,1) \end{matrix}  
                           \bigg| \begin{matrix} -\\
                                       (\frac{u_{_L}}{2},1); (\frac{-u_{_L}}{2},1)  \end{matrix}             
\right)\right.\right.-\\
&~~~~~~~~~~~~~~~~~~~~~~~\left.\left.\frac{\beta_{_{_{E}}}^{\alpha_{_E}}}{\beta_{_{_{E}}}^{\frac{u_E}{2}}\Gamma(\alpha_{_{_{E}}})}
 H_{0,1;1,1;0,2}^{1,0;1,1;2,0} 
\left(  \frac{1}{\beta_{_{_{R}}}},\beta_{_{_{E}}}l
                          \bigg| \begin{matrix} -(\frac{-u_{_E}}{2}+j);(1,1)\\
                                       (-;-) \end{matrix}
                          \bigg| \begin{matrix} (0,1)\\
                                       (0,1) \end{matrix}  
                           \bigg| \begin{matrix} -\\
                                       (\frac{u_{_E}}{2},1); (\frac{-u_{_E}}{2},1)  \end{matrix}             
\right)\right]\right\}
\end{aligned}
\end{equation}
\sepline
\end{figure*}
 \section{Numerical Results and Discussions}
  \begin{figure}[H]
\includegraphics[ height=65mm, width=0.95\linewidth]{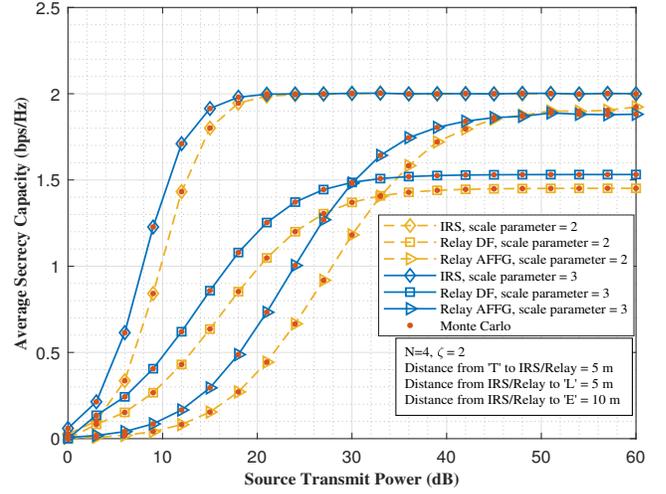}
\caption{ Comparison of average  secrecy  capacity between IRS, DF Relay, and  AFFG Relay in terms of the source transmit power. }
\end{figure}
 We start our numerical results investigation by comparing the average secrecy capacity between IRS, DF, and AFFG for different values of transmitted power $P_{s}$. For the first scenario where we fixed the scale parameters at 2 for all hop-links,  we remark that for the three system models, the secrecy outcomes are very close for $P_{S}<$4 dB. For higher power, we notice that the secrecy capacity margins between the IRS and the two types of relays. For $P_{S}$ = 20 dB, the average secrecy, for IRS, is 2 bps/Hz, which makes it double the output for the AFFG and more than 4 times the DF. This gap, decreases as the power increases, but the IRS still has better results. In the second scenario, we considered the availability of more line-of-sight by increasing the scale parameters to 3. Under this new assumption, we observe that the average secrecy provided by the IRS reaches its ceiling much faster than the relaying systems. Now, we will turn our attention to examine and explain the DF vs. AFFG behavior in the region \RN{2} (see Fig. 4), where the AFFG provides better results than the DF. As we know in the literature, DF always beats AFFG in terms of ergodic capacity.
However, in our study, we are comparing the average secrecy capacity and not the ergodic capacity. For more explanation,  we can observe Fig. 4 which will describe this behavior with accurate calculation. The figure clearly illustrates that in terms of ergodic capacity, DF is better than AFFG. However, the average secrecy capacity has 2 different regions. In region \RN{1}, DF has the higher performance. As an example, we choose  $P_{S}$ = 24 dB (refer to Table. 1). At this point, the average secrecy is 1.2 bps/Hz for DF while it is 0.9 bps/Hz for AFFG. After the point corresponding to $P_{S}$ = 33 dB, DF still better in terms of ergodic capacity at $L$ and $E$, but the difference between those two receiving ends is 1.3 bps/Hz which is less than 1.9 bps/Hz for the AFFG.
 \begin{figure}[H]
\includegraphics[ height=65mm, width=0.95\linewidth]{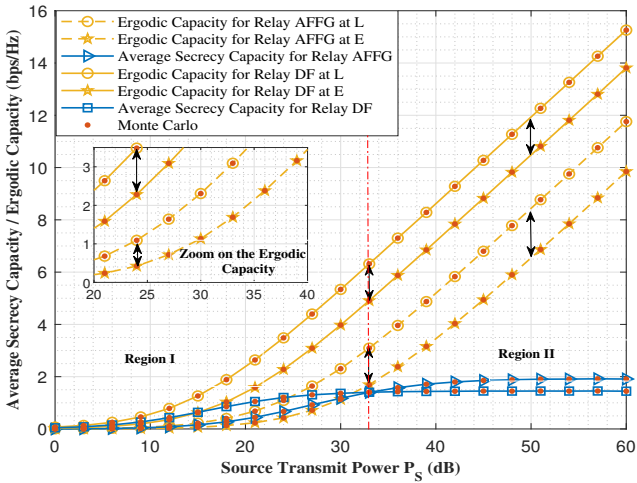}
\caption{Explanation of Relay DF vs. Relay AFFG characteristics in Fig. 3, where the scale parameters = 2.  }
\end{figure}
\begin{table}[H]
\begin{center}
\caption{Selected points from Fig. 4. } 
\begin{tabular}{|c|c|c|c|c|}
\hline
\rowcolor[HTML]{FFCE93} 
\multicolumn{2}{|c|}{\cellcolor[HTML]{FFCE93}Relay Capacity (bps/Hz)}                   & 24 dB & 33 dB & 50 dB \\ \hline
\cellcolor[HTML]{EFEFEF}                       & Ergodic C. at L      & 3.5   & 6.3   & 11.8  \\ \cline{2-5} 
\cellcolor[HTML]{EFEFEF}                       & Ergodic C. at E      & 2.3   & 4.9   & 10.5  \\ \cline{2-5} 
\rowcolor[HTML]{EFEFEF} 
\multirow{-3}{*}{\cellcolor[HTML]{EFEFEF}DF}   & Av. Secrecy Capacity & 1.2   & 1.4   & 1.3   \\ \hline
\cellcolor[HTML]{EFEFEF}                       & Ergodic C. at L      & 1.1   & 3.1   & 8.4   \\ \cline{2-5} 
\cellcolor[HTML]{EFEFEF}                       & Ergodic C. at E      & 0.2   & 1.7   & 6.5   \\ \cline{2-5} 
\rowcolor[HTML]{EFEFEF} 
\multirow{-3}{*}{\cellcolor[HTML]{EFEFEF}AFFG} & Av. Secrecy Capacity & 0.9   & 1.4   & 1.9   \\ \hline
\end{tabular}
\end{center}
\end{table}

Fig. 5, investigates and compares the performances of IRS, DF and AFFG relays for different IRS/Relay to eavesdropper distances.  In the first scenario, when $P_{_S}$ = 10 dB, the secrecy capacity is null where the eavesdropper is less than 8 m, this is due to the fact that the eavesdropper is close to the IRS/Relay than the legitimate receiver  (10 m). When the attacker starts to get farther away from IRS/Relay, the average secrecy starts to raise with a different rate for each system configuration. In other words, and as we can observe from the figure, we get a better result with the IRS  than the DF and AFFG relays. We should note that the relay AFFG under the fixed parameters, is not qualified fit to protect the communication link to the very low average secrecy outcomes, as shown, in the zoom figure. When we double the source power to $P_{_S}$ = 20 dB, the system performance begins to evolve. If the attacker is 30 m away from IRS/Relay,  we will achieve an average secrecy of 0.1 bps while using the AFFG. The DF relay provides 1.25 bps/Hz, which is much higher than the AFFG relay.  On the other side, for the same amount of power and for 4 reflecting elements, the IRS achieves average secrecy of 3.3 bps/Hz,  which is significantly higher than the two relaying models.
\begin{figure}
\includegraphics[height=62mm, width=0.95\linewidth]{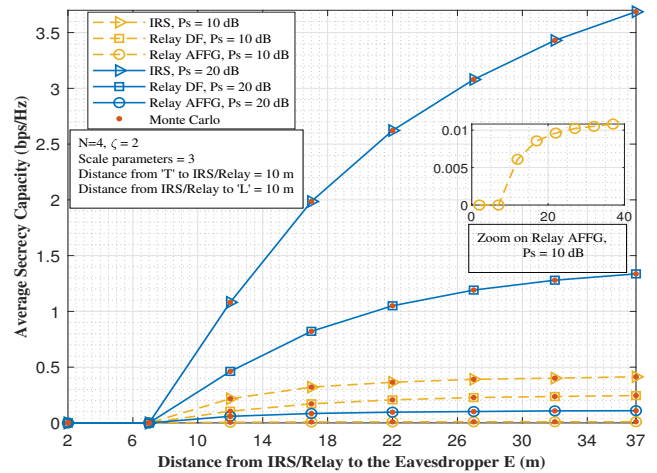}
\caption{ Comparison of average  secrecy  capacity between IRS, DF, and  AFFG relays with respect to the distance between the IRS/Relay to the eavesdropper \textit{E}.  }
\end{figure}

After comparing the IRS to the DF and AFFG relays, we deduce that it recommended to employ the IRS since it offers an acceptable secrecy rate and it outperforms the DF and AFFG relays. Now, we are going to focus on the IRS and study its performances by changing the number of reflecting elements $N$ with respect to the distance from the source to the surface, as shown in Fig. 6. When the BS is very close to the IRS, increasing the number of elements will not significantly improve the average secrecy capacity $C_{_S}^{_{IRS}}$, especially when $N=$ 2 elements. However, as the distance increases, the impact of $N$ becomes more pronounced. For small values of $N$ such as 2 and 8 elements,  $C_{_S}^{_{IRS}}$ decreases with higher rate than for $N$ = 32 and 64 elements as the distance increases. Therefore, As we add more reflecting elements to the IRS, $C_{_S}^{_{IRS}}$ becomes more robust and less sensitive to long distance. Moreover, if we suppose that QoS requires at least a 1.6 bps/Hz for a distance of 15 m and a fixed $P_{_S}$ = 10 dB, we remark that it is not feasible if the IRS has only 2 or 8 reflecting elements. On the other hand, the minimum required threshold is reached when $N$ = 32 elements and more. Therefore, just by increasing N, we can improve the system security with a fixed source power.

 \begin{figure}
\includegraphics[height=62mm, width=0.95\linewidth]{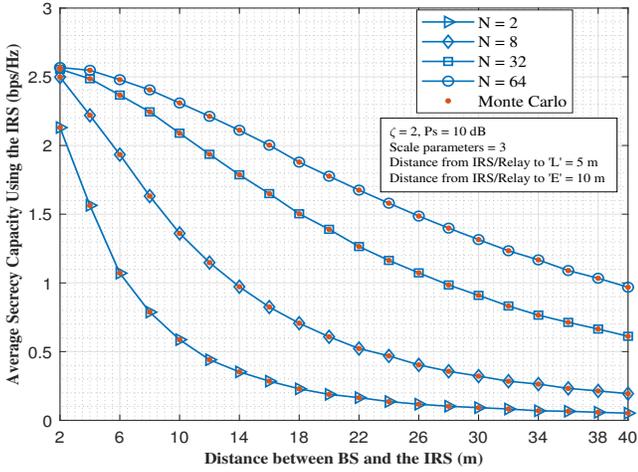}
\caption{ Impact of the number of reflecting elements $N$ on the average secrecy capacity with respect to the distance between the BS and the IRS.  }
\end{figure}

 \section{Conclusion}
In this paper,  we studied the physical layer security in the wireless vehicular network.  We investigated the average secrecy capacity for DF, AFDF relays, and the IRS under the presence of an eavesdropper.  The results showed that the IRS provides higher security outcomes than both, the DF and the AFFG for all the studied scenarios: varying the distance between the attacker and the IRS/relay, changing the transmit source power from the BS, and the scales parameters of the Gamma-distributed channel of each hop.


\bibliographystyle{IEEEtran}
\bibliography{bibliography}

\begin{thebibliography}{10}
\providecommand{\url}[1]{#1}
\csname url@samestyle\endcsname
\providecommand{\newblock}{\relax}
\providecommand{\bibinfo}[2]{#2}
\providecommand{\BIBentrySTDinterwordspacing}{\spaceskip=0pt\relax}
\providecommand{\BIBentryALTinterwordstretchfactor}{4}
\providecommand{\BIBentryALTinterwordspacing}{\spaceskip=\fontdimen2\font plus
\BIBentryALTinterwordstretchfactor\fontdimen3\font minus
  \fontdimen4\font\relax}
\providecommand{\BIBforeignlanguage}[2]{{%
\expandafter\ifx\csname l@#1\endcsname\relax
\typeout{** WARNING: IEEEtran.bst: No hyphenation pattern has been}%
\typeout{** loaded for the language `#1'. Using the pattern for}%
\typeout{** the default language instead.}%
\else
\language=\csname l@#1\endcsname
\fi
#2}}
\providecommand{\BIBdecl}{\relax}
\BIBdecl

\bibitem{realTime}
A.~H. {Khosroshahi}, P.~{Keshavarzi}, Z.~D. {KoozehKanani}, and J.~{Sobhi},
  ``Acquiring real time traffic information using vanet and dynamic route
  guidance,'' in \emph{2011 IEEE 2nd International Conference on Computing,
  Control and Industrial Engineering}, vol.~1, 2011, pp. 9--13.

\bibitem{MeScore}
N.~{Mensi}, A.~{Makhlouf}, and M.~{Guizani}, ``Incentives for safe driving in
  vanet,'' in \emph{2016 4th International Conference on Control Engineering
  Information Technology (CEIT)}, 2016, pp. 1--6.

\bibitem{MeTrafficJam}
N.~{Mensi}, M.~{Guizani}, and A.~{Makhlouf}, ``Study of vehicular cloud during
  traffic congestion,'' in \emph{2016 4th International Conference on Control
  Engineering Information Technology (CEIT)}, 2016, pp. 1--6.

\bibitem{eVCloud}
Y.~{Maalej}, A.~{Abderrahim}, M.~{Guizani}, B.~{Hamdaoui}, and E.~{Balti},
  ``Advanced activity-aware multi-channel operations1609.4 in vanets for
  vehicular clouds,'' in \emph{2016 IEEE Global Communications Conference
  (GLOBECOM)}, 2016, pp. 1--6.

\bibitem{eJEns_Mmwave}
E.~{Balti} and B.~K. {Johnson}, ``Tractable approach to mmwaves cellular
  analysis with fso backhauling under feedback delay and hardware
  limitations,'' \emph{IEEE Transactions on Wireless Communications}, vol.~19,
  no.~1, pp. 410--422, 2020.

\bibitem{mmw}
Y.~{Yang}, Z.~{Gao}, Y.~{Ma}, B.~{Cao}, and D.~{He}, ``Machine learning
  enabling analog beam selection for concurrent transmissions in
  millimeter-wave v2v communications,'' \emph{IEEE Transactions on Vehicular
  Technology}, vol.~69, no.~8, pp. 9185--9189, 2020.

\bibitem{mmwArx}
E.~Balti, N.~Mensi, and S.~Yan, ``A modified zero-forcing max-power design for
  hybrid beamforming full-duplex systems,'' 2020.

\bibitem{SecOut}
H.~{Lei}, H.~{Zhang}, I.~S. {Ansari}, G.~{Pan}, and K.~A. {Qaraqe}, ``Secrecy
  outage analysis for simo underlay cognitive radio networks over generalized-
  $k$ fading channels,'' \emph{IEEE Signal Processing Letters}, vol.~23, no.~8,
  pp. 1106--1110, 2016.

\bibitem{PLS_RIS}
A.~U. Makarfi, K.~M. Rabie, O.~Kaiwartya, X.~Li, and R.~Kharel, ``Physical
  layer security in vehicular networks with reconfigurable intelligent
  surfaces,'' 2019.

\bibitem{RIS_6G2}
C.~Huang, S.~Hu, G.~C. Alexandropoulos, A.~Zappone, C.~Yuen, R.~Zhang, M.~D.
  Renzo, and M.~Debbah, ``Holographic mimo surfaces for 6g wireless networks:
  Opportunities, challenges, and trends,'' \emph{ArXiv}, vol. abs/1911.12296,
  2019.

\bibitem{RIS_6G1}
I.~Yildirim, A.~Uyrus, E.~Basar, and I.~F. Akyildiz, ``Propagation modeling and
  analysis of reconfigurable intelligent surfaces for indoor and outdoor
  applications in 6g wireless systems,'' \emph{ArXiv}, vol. abs/1912.07350,
  2019.

\bibitem{IRSvsRelayEmil}
E.~{Bj\"{o}rnson}, o.~{\"{o}zdogan}, and E.~G. {Larsson}, ``Intelligent
  reflecting surface versus decode-and-forward: How large surfaces are needed
  to beat relaying?'' \emph{IEEE Wireless Communications Letters}, vol.~9,
  no.~2, pp. 244--248, 2020.

\bibitem{IRSvsRelay}
M.~{Di Renzo}, K.~{Ntontin}, J.~{Song}, F.~H. {Danufane}, X.~{Qian},
  F.~{Lazarakis}, J.~{De Rosny}, D.~{Phan-Huy}, O.~{Simeone}, R.~{Zhang},
  M.~{Debbah}, G.~{Lerosey}, M.~{Fink}, S.~{Tretyakov}, and S.~{Shamai},
  ``Reconfigurable intelligent surfaces vs. relaying: Differences,
  similarities, and performance comparison,'' \emph{IEEE Open Journal of the
  Communications Society}, vol.~1, pp. 798--807, 2020.

\bibitem{MRC}
K.~A. {Hamdi}, ``Capacity of mrc on correlated rician fading channels,''
  \emph{IEEE Transactions on Communications}, vol.~56, no.~5, pp. 708--711,
  2008.

\bibitem{BesselToMeijer}
V.~S. Adamchik and O.~I. Marichev, ``The algorithm for calculating integrals of
  hypergeometric type functions and its realization in reduce system,''
  \emph{Proceedings of the international symposium on Symbolic and algebraic
  computation - ISSAC 90}, 1990.

\bibitem{Tab}
I.~S. Gradshteyn and I.~M. Ryzhik, \emph{Table of integrals, series, and
  products}, 7th~ed.\hskip 1em plus 0.5em minus 0.4em\relax The
  Netherlands:Academic, 2007.

\bibitem{eAppr}
E.~{Balti}, M.~{Guizani}, and B.~{Hamdaoui}, ``Hybrid rayleigh and
  double-weibull over impaired rf/fso system with outdated csi,'' in \emph{2017
  IEEE International Conference on Communications (ICC)}, 2017, pp. 1--6.

\bibitem{eTransG}
E.~{Balti} and M.~{Guizani}, ``Mixed rf/fso cooperative relaying systems with
  co-channel interference,'' \emph{IEEE Transactions on Communications},
  vol.~66, no.~9, pp. 4014--4027, 2018.

\bibitem{eRelayFSO2}
E.~{Balti}, M.{Guizani}, B.~{Hamdaoui}, and Y.~{Maalej}, ``Partial relay
  selection for hybrid rf/fso systems with hardware impairments,'' in
  \emph{2016 IEEE Global Communications Conference (GLOBECOM)}, 2016, pp. 1--6.

\bibitem{eRelayFSO3}
E.~{Balti}, M.~{Guizani}, B.~{Hamdaoui}, and B.~{Khalfi}, ``Aggregate hardware
  impairments over mixed rf/fso relaying systems with outdated csi,''
  \emph{IEEE Transactions on Communications}, vol.~66, no.~3, pp. 1110--1123,
  2018.

\bibitem{eRelayFSO1}
------, ``Mixed rf/fso relaying systems with hardware impairments,'' in
  \emph{GLOBECOM 2017 - 2017 IEEE Global Communications Conference}, 2017, pp.
  1--6.

\bibitem{TripFoxInt}
P.~K. Mittal and K.~C. Gupta, ``{An integral involving generalized function of
  two variables},'' \emph{Proceedings of the Indian Academy of Sciences -
  Section A}, vol.~75, no.~3, pp. 117--123, 1972.

\bibitem{eT}
E.~Balti, ``\BIBforeignlanguage{English}{Analysis of hybrid free space optics
  and radio frequency cooperative relaying systems},'' Master's thesis, 2018.

\bibitem{eBivariate_Relay}
E.~{Balti} and M.~{Guizani}, ``Impact of non-linear high-power amplifiers on
  cooperative relaying systems,'' \emph{IEEE Transactions on Communications},
  vol.~65, no.~10, pp. 4163--4175, 2017.

\end{thebibliography}
\end{document}